\documentclass[crcready,onecolumn]{iosart2x}

\makeatletter
\let\numberlines@hook\relax
\makeatother

\usepackage{dcolumn}

\pubyear{0000}
\volume{0}
\firstpage{1}
\lastpage{1}

\begin{document}
\begin{frontmatter} 

\title{Development of High Intensity Neutron Source at the European Spallation Source}

\runningtitle{Development of High Intensity Neutron Source at the European Spallation Source}

\author[A,B]{\inits{F.}\fnms{V.} \snm{Santoro}\ead[label=e1]{valentina.santoro$@$esss.se}%
\thanks{Corresponding author. \printead{e1}.}},%
\author[A]{\inits{T.}\fnms{K.H.~}\snm{Andersen}\ead[label=e3]{third@somewhere.com}},
\author[A,B]{\inits{T.}\fnms{ D.~D.~}\snm{DiJulio}\ead[label=e3]{third@somewhere.com}},
\author[C]{\inits{T.}\fnms{E.~B.~}\snm{Klinkby}\ead[label=e3]{third@somewhere.com}},
\author[A]{\inits{T.}\fnms{T.~M.~}\snm{Miller}\ead[label=e3]{third@somewhere.com}},
\author[D]{\inits{T.}\fnms{D.~}\snm{Milstead}\ead[label=e3]{third@somewhere.com}},
\author[A]{\inits{T.}\fnms{G.~}\snm{Muhrer}\ead[label=e3]{third@somewhere.com}},
\author[E]{\inits{T.}\fnms{M.~}\snm{Strobl}\ead[label=e3]{third@somewhere.com}},
\author[A]{\inits{T.}\fnms{A.~}\snm{ Takibayev}\ead[label=e3]{third@somewhere.com}},
\author[A]{\inits{S.}\fnms{L.} \snm{Zanini}\ead[label=e2]{second@somewhere.com}},
and
\author[F]{\inits{T.}\fnms{O.~}\snm{Zimmer}\ead[label=e3]{third@somewhere.com}}

\runningauthor{V.~Santoro et al.}
\address[A]{European Spallation Source ERIC, P.O. Box 176, 22100 Lund, Sweden}
\address[B]{Department of Physics, Lund University, 22100 Lund, Sweden} 
\address[C]{DTU Nutech, \orgname{Technical University of Denmark, DTU Ris{\o} Campus, Frederiksborgvej 399, DK-4000, Roskilde, Denmark,}}
\address[D]{Stockholm University, SE-106 91 Stockholm, Sweden }
\address[E]{Paul Scherrer Institut, PSI Aarebr\"Ucke, 5232 Villigen, Switzerland}
\address[F]{Institut Laue-Langevin - 71 avenue des Martyrs CS 20156, 38042 GRENOBLE Cedex 9 - France}

\begin{abstract}
The European Spallation Source being constructed in Lund, Sweden will provide the user community with a neutron source
of unprecedented brightness. By 2025, a suite of 15 instruments will be served by a high-brightness moderator system
placed above the spallation target. The ESS infrastructure, consisting of the proton linac, the target station, and the
instrument halls, allows for implementation of a second source below the spallation target. We propose to develop a second
neutron source with a high-intensity moderator able to (1) deliver a larger total cold neutron flux, (2) provide high intensities
at longer wavelengths in the spectral regions of Cold (4-10 \AA ), Very Cold (10-40 \AA ), and Ultra Cold (several 100 \AA ) neutrons,
as opposed to Thermal and Cold neutrons delivered by the top moderator. Offering both unprecedented brilliance, flux, and
spectral range in a single facility, this upgrade will make ESS the most versatile neutron source in the world and will further
strengthen the leadership of Europe in neutron science. The new source will boost several areas of condensed matter
research such as imaging and spin-echo, and will provide outstanding opportunities in fundamental physics investigations of
the laws of nature at a precision unattainable anywhere else. At the heart of the proposed system is a volumetric liquid
deuterium moderator. Based on proven technology, its performance will be optimized in a detailed engineering study. This
moderator will be complemented by secondary sources to provide intense beams of Very- and Ultra-Cold Neutrons. 
\end{abstract}

\end{frontmatter}

\section{Introduction}
Presently under construction, the European Spallation Source (ESS) in Lund, Sweden, is a multi-disciplinary
international laboratory with 13 European member states and will be one of Europe's, flagship facilities. It will
operate the world's most powerful pulsed neutron source. Initially, the spallation source will be equipped with
only a single compact low-dimensional moderator~\cite{1}, which has been designed to deliver brightest neutron beams
for condensed matter experiments, optimized for small samples, flexibility and parametric studies. A user
program will start in 2023, and by 2025 a suite of 15 neutron scattering instruments will be installed at this first
neutron source. The flexible design of ESS, however, leaves a great opportunity to implement a second source
with complementary characteristics going well beyond the initial goals of the facility development~\cite{2}. This new
infrastructure must be capable of delivering highest possible total intensity (as opposed to brightness which was
the main design criterion for the first source) of cold neutrons with wavelengths above 4 \AA, including neutrons
in the long-wavelength tails of the spectrum, so-called Very Cold (VCN) and Ultra Cold Neutrons (UCN). 
The new infrastructure enables experiments in two main
categories: condensed matter research using techniques like spin-echo, SANS and imaging, and fundamental
physics (searches for neutron-antineutron oscillation, a non-zero neutron electric dipole moment, a fifth force,
extra dimensions etc.). The new source will enable a program of world-class experiments, often with a unique
physics reach. The sensitivity on various measurements and searches can be improved by up to three orders of
magnitude compared with best running facilities.
An alternative source with comparable parameters to enable these experiments does not exist. There are two
other high-power, MW-class, pulsed spallation sources in the world: J-PARC in Japan, and SNS in USA.
Upgrade plans exist for implementation of second target stations at both facilities \cite{jparcupgrade,snsupgrade} , and in both cases the focus
is on using low-dimensional high-brightness moderators, as for the ESS first source. These upgrades aim to
deliver superior peak brightness to the ESS first source (up to a factor of 5), but with a time-average brightness
which remains about a factor of 5 lower than the ESS first source, when operating at 5 MW. The combined
effect is that after these upgrades, the SNS and J-PARC second target stations will deliver performance which
is similar to that delivered by the ESS first source, and with a similar spectral emphasis on the thermal and cold
neutron ranges.
An intense source at ESS, focused on delivering more neutrons with longer wavelengths, would outperform
these planned upgrades in the long-wavelength regime by more than an order of magnitude. 
The installation of a long-wavelength facility at the ESS is uniquely favorable due to the high proton
power and long-pulse time structure. No existing facilities can upgrade in this way.

\section{The ESS source upgrade }
Neutrons are a valuable research tool for the investigation of matter in a variety of fields.  The application of neutrons as a research probe started in the 50's with the advent of research reactors. Developments of intense neutron sources followed two threads, leading to high-flux reactors, such as at ILL, and high-power spallation sources such as at SNS, J-PARC and now ESS. Despite these developments, neutron scattering remains an intensity-limited technique, due to the fact that only a very small fraction of the neutrons produced reaches a sample or an experiment. The ESS has developed high-brightness, low-dimensional moderators to maximize the availability of neutrons for typical experiments that are possible only at ESS, in particular for small samples with low-divergence beams.  This optimization has led to a single moderator system capable of satisfying the needs of all the ESS instruments in the initial suite. Many other experiments and applications would however strongly profit from a source with different, complementary characteristics, notably those requiring a larger total number of neutrons emitted from the source. This latter feature is however crucial for many fundamental physics experiments, such as the proposed neutron-antineutron oscillation search NNBAR~\cite{nnbar}. Also neutron scattering experiments would benefit, investigating a sample with techniques allowing for large beam divergence. Moreover, some experiments can strongly profit from using neutrons with longer wavelengths than delivered by a cold hydrogen moderator (2-10 \AA ). In particular, the range 10-40 \AA, which we can broadly call Very Cold Neutrons (VCNs), is of great interest for some neutron scattering applications (spin-echo, SANS), neutron imaging, and notably for fundamental physics (NNBAR, searches for a fifth force and extra dimensions, etc.). The combination of two distinctive features of a new cold source that will be developed, i.e., the availability of a higher total intensity and a colder spectrum with much enhanced output at longer neutron wavelengths, is extremely attractive for the communities of neutron users.
The following Figures  \ref{fig1},~\ref{fig2},~\ref{fig5}  show the arrangement of the present moderator system at ESS, and areas still available for additional future instruments that can be fed by a second moderator system. These free areas are marked in green areas in Fig.~\ref{fig5} they represent a valuable infrastructure for future developments.
 
\begin{figure}[t]{
\includegraphics[width=0.56\linewidth, angle=0]{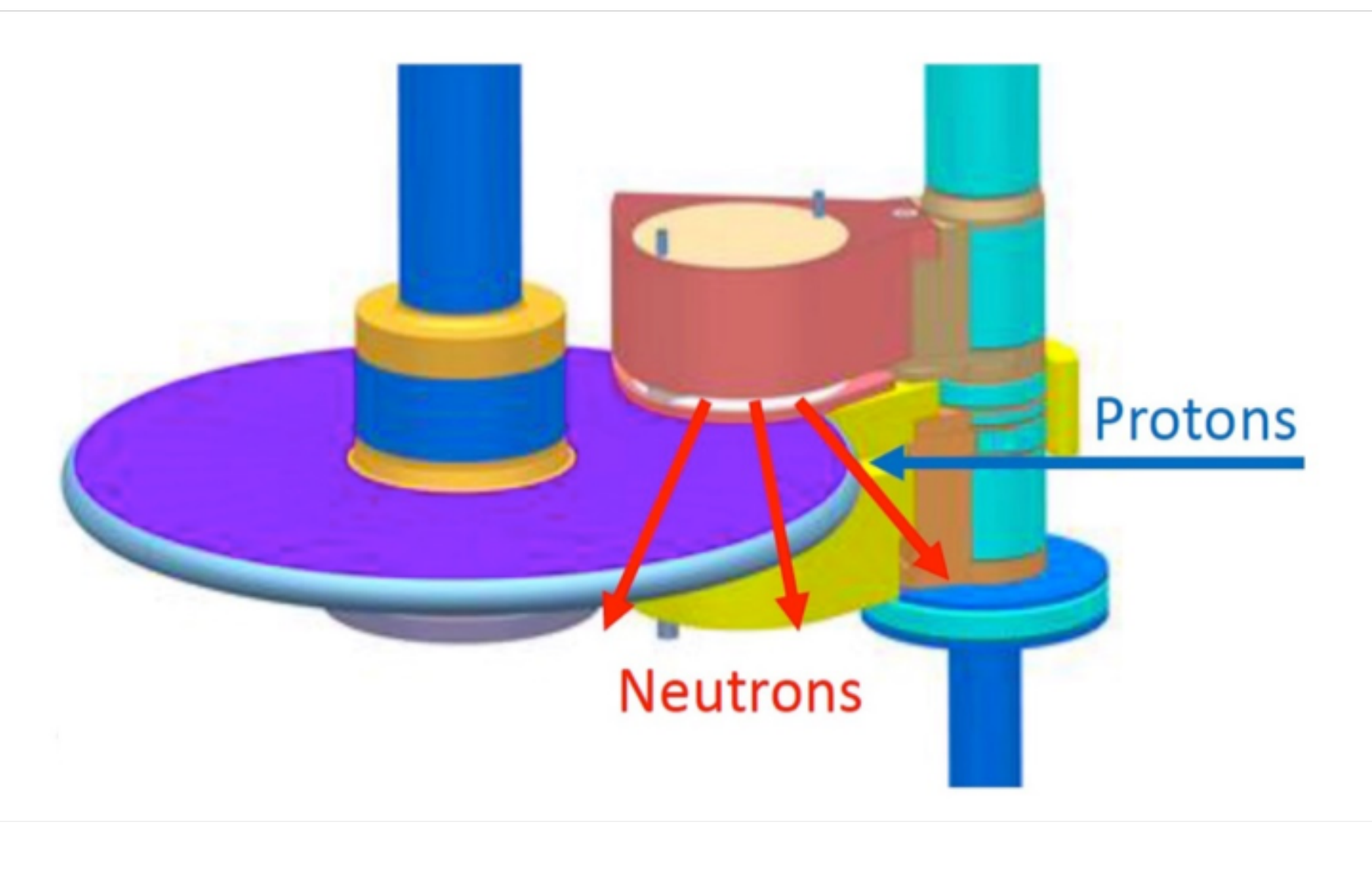}}
\caption{ Schematic view of the ESS target-moderator-reflector system. Protons with 2 GeV energy impinge on a rotating target consisting of tungsten (purple target in the figure). A cylindrical steel structure (dark red) placed above the target contains the moderator and reflector from which neutrons are extracted to the beam lines (red arrows). A similar container (yellow) is placed below the target. Presently unused, it can host another moderator-reflector system (compare Fig.\ref{fig2})}
  \label{fig1}
\end{figure}
 
 \begin{figure}[t]{
\includegraphics[width=0.56\linewidth, angle=0]{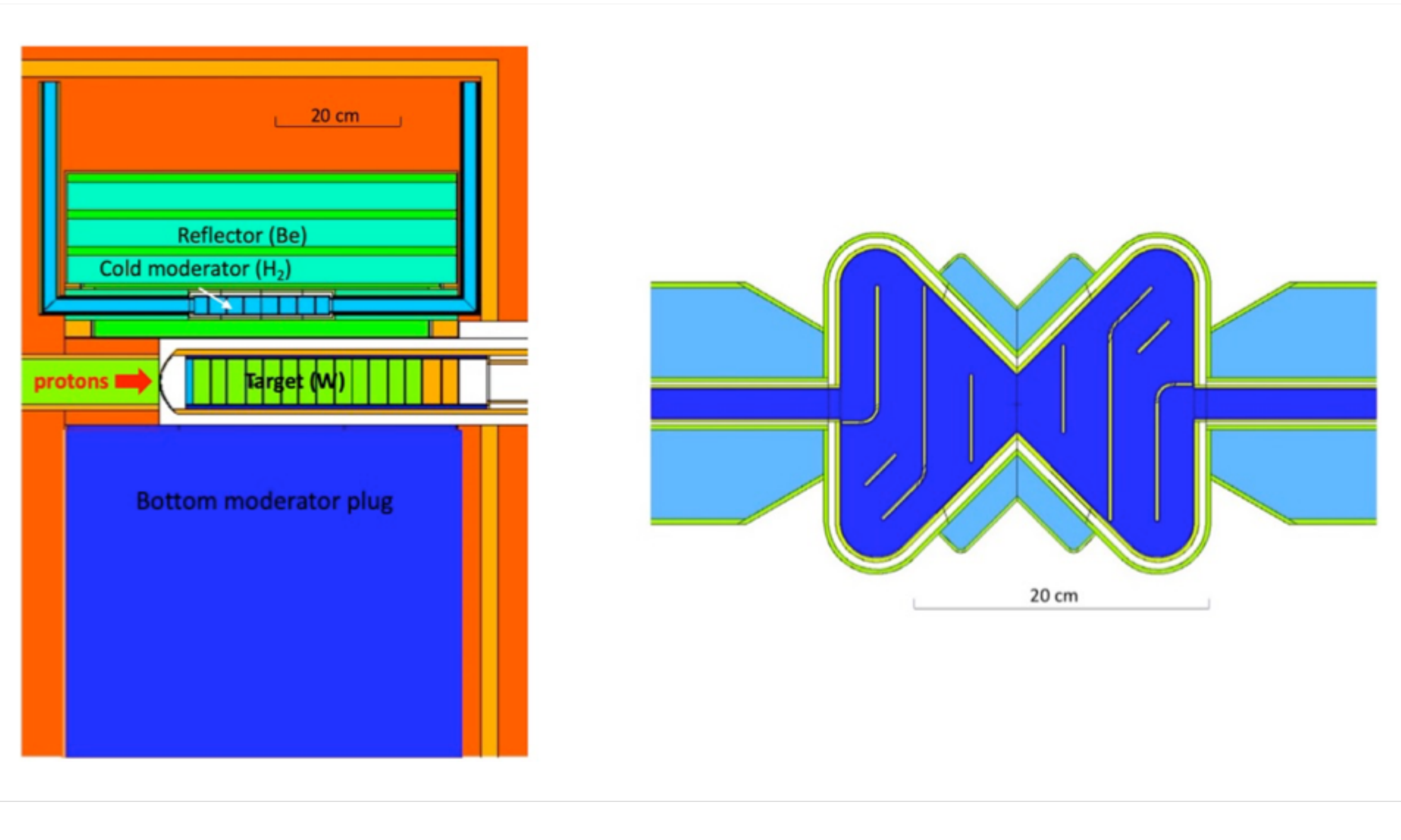}}
\caption{Left: MCNP model of the present configuration of the ESS target-moderator-reflector. A high-brightness moderator system (right figure) is placed above the spallation target. The place below the target is presently occupied by a steel plug and can accommodate a second moderator system}
  \label{fig2}
\end{figure}

 \begin{figure}[t]{
\includegraphics[width=0.67\linewidth, angle=0]{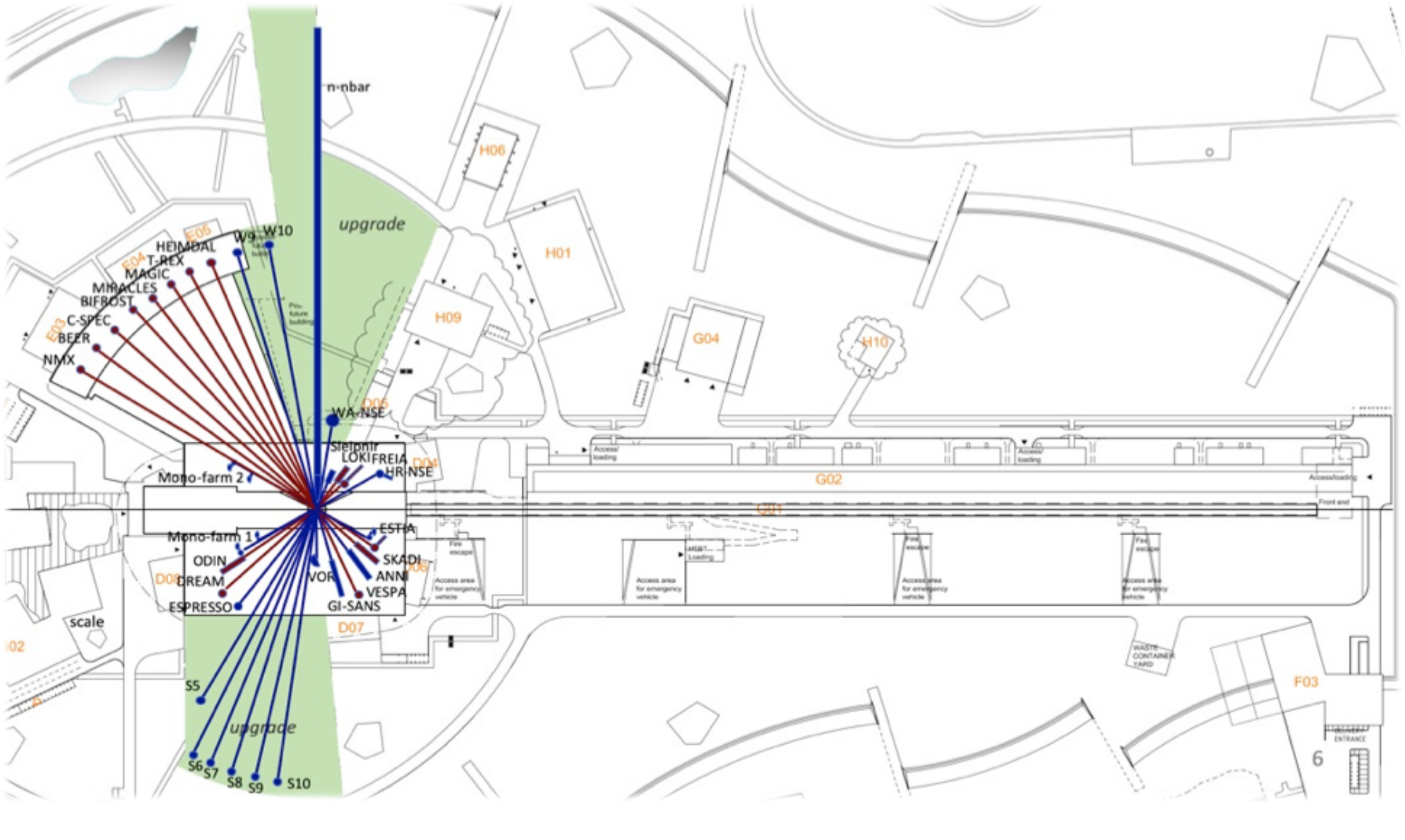}}
\caption{Possible upgrade plan of ESS: the part marked in green covers experimental areas of the facility which are left free by the initial suite of 15 instruments.}
  \label{fig5}
\end{figure}

\begin{figure}[t]{
\includegraphics[width=0.49\linewidth, angle=0]{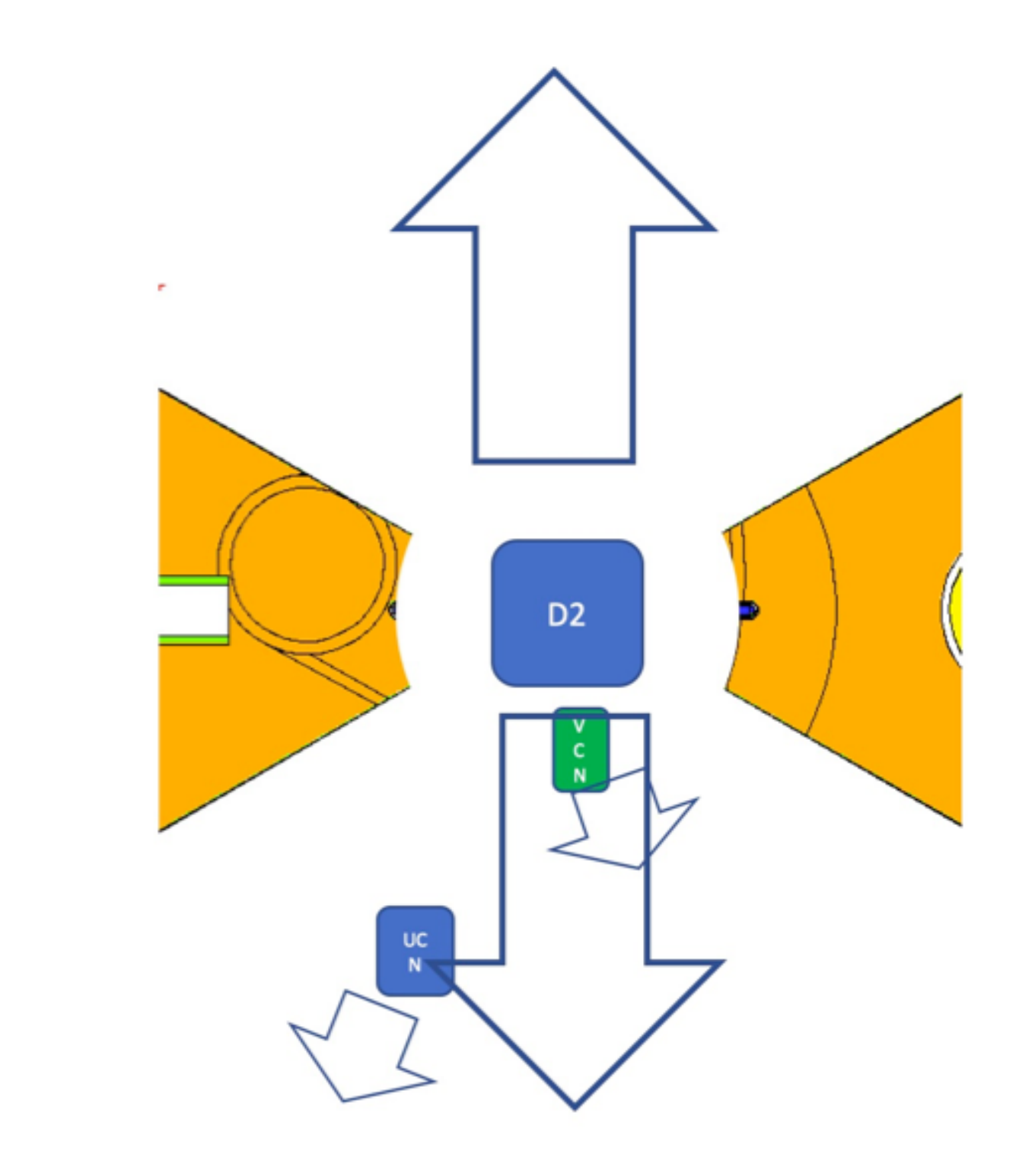}}
\caption{Conceptual arrangement of a possible moderator configuration below the target, see explanations in the text.}
  \label{fig6}
\end{figure}

It is important to note the difference between the brightness and the total intensity of a neutron source. The latter is given by the total flux of neutrons emitted from the surface, whereas the source brightness is the flux density in a neutron phase space element of interest for use in a neutron instrument. For some applications, notably if small samples are investigated in neutron scattering experiments, a small moderator with higher brightness is more interesting than a large moderator with higher total flux but smaller brightness. It has been shown \cite{2} that an increase of the size of a para-hydrogen moderator from the 3 cm foreseen for the upper ESS moderator to 10 cm would lead only to a gain of 20\% in total intensity. While the instruments viewing the upper hydrogen moderator can profit from its high brightness, other applications described in section 
\ref{applications} are able to take advantage of the higher total intensity offered by a large LD$_{2}$ moderator (for which the argument of trading brightness versus intensity does not hold due to the different neutron scattering physics it involves).
As shown schematically in Fig.~\ref{fig6}, we propose therefore to place in the central position below the target a second cold source containing approximately 20 liters of LD$_{2}$. This type of moderator seems currently the best choice for high-intensity applications. It is technically state of the art and well established as user facilities at ILL and SINQ at PSI \cite{sinq} . With respect to the upper para-hydrogen moderator, which was designed as a bi-spectral moderator with high-brightness, the LD$_{2}$ moderator will provide a colder neutron spectrum. Depending on its exact location, an about 3-4 times higher total intensity has been estimated in a preliminary calculation . Neutrons emitted from a surface area of about 20$\times$20 cm$^{2}$ can be either used directly for a variety of applications (see section \ref {applications}), or be further cooled down to the VCN or even UCN energy ranges in secondary cold sources to be placed in close vicinity to the LD$_{2}$ moderator, as discussed in the next sections. The LD$_{2}$ moderator will thus serve the two purposes to
\begin{itemize}
\item	Directly feed beamlines, possibly with an advanced reflector system to enhance the CN and VCN fluxes,
\item Feed a UCN source, and eventually a VCN moderator, for generation of high fluxes of UCNs and VCNs, respectively.
\end{itemize}

The intensity of the new LD$_{2}$ moderator for ESS can be maximized in two ways, 
\begin{itemize}
\item by using optimized geometries. One may also include reentrant holes, as used in several facilities for cold and thermal moderators, to increase the flux in specific emission channels,
\item by using advanced reflector materials optimized for the CN and VCN wavelength ranges. These are expected to further increase the intensity gain. 
\end{itemize} 
 While the first possibility involves optimization of well-established technology in the context of the special conditions at ESS, the second is more explorative and requires extensive studies of suitable advanced materials. The study of this advanced materials is part of the upgrade plan of ESS.

\subsection{Source for ultra-cold neutrons}
Several options for sources of ultra-cold neutrons (UCN) at spallation sources have been discussed at the ESS Science Symposium on Neutron Particle Physics at Long Pulse Spallation Sources, NPPatLPS, held in 2013 at LPSC in Grenoble  \cite{3,4,5,6}. They commonly use superfluid helium (He-II) as a medium for conversion of meV cold neutrons to the UCN energy range of typically 100 neV. UCN production in a He-II converter relies on the possibility of nearly entire neutron energy loss in single scattering events, mostly via a one-phonon process induced by neutrons of 8.9~\AA wavelength \cite{7} . Sources can be classified as ``in-beam"  and ``in-pile", depending on the distance to the primary moderator. The in-beam option places a He-II converter at the end of a neutron guide, as initially proposed in Ref.~\cite{8} . This is viable due to the strictly vanishing absorption cross section of $^{4}$He. A user facility of this type, called SuperSUN, is currently being commissioned at the ILL. Its position remote from the harsh radiation environment enables converter temperatures below 0.6 K where phonon up-scattering becomes so small that the storage lifetime in the converter is dominated by neutron beta decay. In addition, a magnetic reflector can be implemented suppress UCN losses at the converter walls. These two aspects together lead to projected UCN densities in the source beyond 1000 cm$^{-3}$, as compared to typically few 10 cm$^{-3}$~in the currently best UCN sources for users. At the ESS, UCN densities of an in-beam source might reach the order of 10$^{4}$ cm$^{-3}$ for an efficient cold-neutron extraction under large solid angle attainable with modern supermirror technology. To be noted is that an in-beam source attains its high UCN density only after a saturating period of neutron accumulation, which becomes then diluted on UCN delivery to an actual experiment. This type of source is therefore reasonably applied only for experiments involving small chambers for UCN storage. An ``in-pile" UCN source at a spallation source is located in the intense field of cold neutrons generated by a primary cold source situated close to the spallation target. Since the UCN production rate is proportional to the flux of incident neutrons, it is typically by two to three orders of magnitude larger in an in-pile source than at an external beam. However, this requires removal of considerable heat loads on the converter due to radiation, which excludes operation of the source below 1 K. Since the UCN storage lifetime due to phonon up-scattering in this temperature range decreases with the 7th power of the converter temperature, the gain in UCN density does not fully follow the gain in UCN production. A significant benefit of the much higher UCN production rate in the more powerful in-pile sources is a much weaker dependence of UCN density on chamber size. They thus combine high density and high flux and are therefore perfectly suited to fill larger chambers, and for use in non-storage, flight-through experiments. 
A dedicated  engineering analysis will be carried out to study the best possibility for an in-pile UCN source at ESS. Among several UCN source projects worldwide, proposals for in-pile He-II UCN sources for the PIK reactor~\cite{serebrov} at Gatchina, Russia, are of particular interest for the ESS. They promise UCN densities in the range of a few 10$^{3}$ to~10$^{4}$~cm$^{-3}$ and are partly well advanced so that some of their concepts can be taken as a starting point for our analysis.

\subsection{Source for very cold neutrons}
The availability of intense fluxes of Very Cold Neutrons is potentially a game changer in several neutron scattering applications, as well as in fundamental physics research with neutrons. The positive impact of longer neutron wavelengths $\lambda$ on the performance of various classes of instruments can be seen from the $\lambda$ dependences of the instrumental resolution at fixed geometry, and the intensity at fixed resolution, e.g., $\lambda^{-1}$ and $\lambda^{2}$ for reflectometers, $\lambda^{-3}$ and $\lambda^{2}$for Time-of-Flight (ToF) instruments, or $\lambda^{-3}$ and $\lambda^{3}$ for neutron spin echo~\cite{12}.

Particle-physics experiments able to take advantage of higher VCN fluxes are the search for neutron anti-neutron oscillations (see Section~\ref{applications}) for which the discovery potential is proportional to $\lambda^{2}$, a project of an in-beam search for a non-vanishing neutron electric dipole moment~\cite{piegsa1}, and experiments searching for new fundamental forces~\cite{piegsa2}.
We will study two possibilities to get these intense fluxes,
\begin{itemize} 
\item with a dedicated VCN moderator  using a novel material 
\item using VCN extraction from the colder tail of the LD$_{2}$ moderator spectrum, by enhancing the spectral part beyond 10 \AA~using advanced reflectors. 
\end{itemize}
For the first option, we have to perform dedicated studies to this novel material (see Section.~\ref{reflectors}) since thre is still  lacking of a a detailed knowledge of important properties of materials identified as suitable for moderation to the VCN energy range.
For the second option, we intend to use cold-neutron reflectors to enhance the VCN yield from the LD$_{2}$ source. 

\subsection{Advanced Reflector materials} 
\label{reflectors}

Recent years have seen intense research towards novel moderator/reflector materials and for beam extraction systems, with the aim to improve cold-neutron sources. The following materials were identified as promising candidates:
\begin{itemize}
\item {\bf Diamond nanoparticles (NDs)} have been studied in the last 15 years and theoretical and experimental results indicate that these materials have large albedo for cold and sub-cold neutrons. There are now studies ongoing or planned at most of the major neutron scattering facilities worldwide, including SNS, JPARC, ISIS, ANSTO, and ILL. One of the reasons for the growing interest is a recent measurement of the scattering cross section in the CN/VCN regime~\cite{nano1} - which is found to be larger by up to two orders of magnitude than of carbon nuclei without coherent amplification at long wavelengths. These experiments agree well with a simple theoretical picture given in Refs.~\cite{nano2,nano3}. Coherent scattering by NDs notably gives rise to small angle scattering in the CN regime, which becomes almost isotropic in the VCN regime. These properties can be exploited for cold neutron sources. 
The first property, i.e., quasi-specular reflection of cold neutrons might enable neutron transport in a diffusive channel coupled to a moderator. It is to be shown how effective this mechanism is with respect to neutron optical beam transport by supermirrors. If the inner walls facing the neutron beam in a beam extraction would be made of NDs, stray neutrons leaving the moderator at a too large divergence to be of experimental value, might scatter off the NDs, and get a second chance to be emitted within the acceptance of the optics. This could become useful for transport of cold neutrons to increase the flux at a sample. An important application of a diffusive channel will be the enhancement of CN fluxes from a primary LD$_{2}$ moderator to feed a secondary moderator for VCN or UCN production, placed at some distance, providing a corresponding increase of VCN or UCN fluxes. The second interesting property of NDs, a large albedo for VCNs, suggests to use them as a reflector surrounding a VCN moderator to enhance its performance.\\
\item { \bf MgH$_{2}$} is another promising material which has been proposed for diffuse reflection of cold neutrons~\cite{granada}. 
Preliminary studies have been performed and more detailed characterizations are ongoing.\\ 
\item {\bf Graphite intercalation compounds}  (GICs) have the potential to act as high-performance reflectors for CNs and VCNs. By intercalating atoms such as fluorine or oxygen between the graphene layers, the largest d-spacing can be increased by a factor of two or more in comparison to regular graphite, pushing the Bragg cut-off to 8-10~\AA, or even beyond. A polycrystalline assembly of such a material will be an effective reflector for wavelengths up to that Bragg cut-off, which is to be compared to a cut-off wavelength of about 4~\AA for polycrystalline beryllium, the most commonly-used reflector material for CN sources. A variety of such GIC compounds can be synthesized, and the most promising materials will be tested for their neutronic performance. For such an application, the requirements are: large lattice parameters, high coherent scattering cross section, low absorption cross section, low chemical reactivity, and resistance to radiation damage. \\
\item  {\bf Deuterated clathrate hydrates } have been proposed more recently ~\cite{zimmer} to cover the CN range from standard LD$_{2}$ cold sources, for which they promise untypically large albedos. These inclusion compounds, consisting of a network of water molecules that form cages able to host small guest molecules, are also particularly promising for VCN production, so that a cooled reflector would also possess moderating properties. This is due to their cross section for inelastic incoherent neutron scattering by local modes, which can remove energy from the neutrons without kinematic restrictions due to a dispersion relation.  
\end{itemize}

 However, the characterization of this material class for this application is still at a preliminary stage so additional experimental studies must be performed to be used in the future ESS upgrade. 

\section{High intensity neutron source applications }
\label{applications}

As stated previuosly a high intensity source would benefit several different applications from condensed matter science to fundamental physics. Below a detailed description of the 
scientific impact of such new source on different applications is provided.

\subsection{Condensed -matter science }

\subsubsection{Neutron spin-echo (NSE) }

Neutron spin-echo is the technique which can measure the slowest dynamics accessible with neutrons.
It does this by decoupling the energy resolution from the degree of beam monochromaticity and measures the
van Hove correlation function directly in the time regime, rather than its Fourier transform as is conventionally
recorded in neutron spectroscopy. This makes NSE a unique tool for characterizing the motion of large
molecules and it is very often the only direct test of predictions made by molecular dynamics (MD) simulations.
Significantly higher CN or VCN fluxes increase the performance of NSE spectroscopy in several ways: Firstly,
by increasing the longest relaxation times which the technique can access. Secondly, by allowing measurements
at larger wave vector transfer where the scattering cross sections are lower, probing movements associated with
smaller structures. It will also allow much faster measurements, increasing throughout for a technique which is
strongly flux-limited. Finally, it will permit the study of much smaller amounts of sample, which is extremely
important for biological studies. The expected  flux increase will be a game-changer for spin-echo, making the technique
available to non-expert users and tapping into a much larger user community than currently addressed.

\subsubsection{Small-angle neutron scattering (SANS)  }

SANS measures large structures such as macromolecules in solution by
detecting the scattering at small angles using cold neutrons. Longer wavelength neutrons can be used to access
larger structures, or alternatively, the same length scales can be accessed by measuring at larger scattering
angles, potentially improving the performance of the technique. In addition, the generally lower source flux at
these wavelengths can be compensated for by employing novel focusing techniques, such as magnetic or
materials lenses using refraction whose effectiveness increases rapidly with wavelength.
A significant increase in source intensity of wavelengths in the 5-30 \AA~range could be transformative for this
field. It will allow the measurement of smaller volumes and greater length-scales, overlapping with the lengthscale
regime of neutron imaging. Current efforts to access these length scales, known as V-SANS or U-SANS
(very- or ultra- small-angle neutron scattering) typically suffer from very long counting times and employ
elaborate instrument concepts designed to compensate for the very low source brightness of neutrons in that
wavelength range. Rapid measurements with small beams will also enable scanning SANS measurements which
can probe local structures and thus structural variations throughout inhomogeneous samples, potentially even
in 3D. Faster measurements on the other hand will enable the study of processes and structural transformations
that are not amenable today.
Two instruments currently under construction, viewing the upper moderator are set already today to transform
the field through outstanding performance beyond current capabilities enabling the measurement of smaller
samples or faster kinetics. The gains envisioned here are in addition to those of these current projects. They will
enable measurements of yet even smaller structures and faster processes and provide data with high spatial resolution
through fast pencil beam scanning procedures, today only known from synchrotron sources, but benefitting
from the unique contrast conditions which only neutrons can provide.

\subsubsection{Neutron imaging }

Neutron imaging is a rapidly-evolving real space technique probing macroscopic structures with a very wide
range of applications, covering both academic research and areas with direct industrial impact. The ODIN\cite{instruments}
instrument at ESS will provide an order-of-magnitude improvement in versatility and performance compared to
existing instruments, and is expected to be strongly over-subscribed. An imaging instrument viewing the ESS
lower moderator would provide even further enhanced performance in a complementary regime of applications
with respect to the ODIN instrument, in particular for large industrial samples and high temporal and spatial
resolution applications. The lower moderator position provides a unique opportunity for this by significantly
improving two aspects:
\begin{itemize}
\item Large moderator surface: The upper moderator is optimized for high (local) brightness, rather than high (areaintegrated)
intensity. A moderator with a viewable area of the order of 20$\times$20 cm$^{2}$, compared to the 3$\times$10 cm$^{2}$
area used by ODIN, will allow a much larger and more uniform field of view. This will both extend the
measurement capability to much larger and more complex systems and industrial components or whole
machinery.
\item Long wavelength neutrons: A flux enhancement for neutrons in the 5-20 \AA~ range will greatly increase, e.g., the
sensitivity of polarized neutron imaging, extending the range of accessible 3D magnetic field distributions to
lower fields and local electric currents, such as in energy conversion devices. The long-wavelength regime will
also greatly improve quantitative high resolution imaging with neutrons, as the signal is not affected by the
diffraction background which is otherwise unavoidable from crystalline materials.
\end{itemize}

\subsection{Fundamental Physics}

\subsubsection{The NNBAR project to search for neutron-antineutron oscillations}

Fundamental particle physics research is a  field which has developed
by exploring particle interactions and discovering new particles at progressively higher energy scales. Collider
experiments, such as the CERN Large Hadron Collider (LHC) are generally probe only to mass scales up to the
collision energy. Non-collider experiments study or search for specific processes at an extremely high precision
are sensitive to hitherto unseen physics processes at mass scales substantially in excess of those available
at colliders. Taking advantage of the unique potential of the ESS, the NNBAR experiment will continue the
long tradition of high sensitivity small scale particle physics experiments by performing searches for neutrons
converting to antineutrons. NNBAR has a unique reach in sensitivity to new physics beyond any conceivable
or running collider experiment. This enables the experiment to tackle some of the most important open questions
in modern physics. One such issue is the so-called matter asymmetry of the universe. While the world around
us consists of matter particles (e.g., electron or neutrons) antimatter particles (e.g., positrons or antineutrons)
are typically glimpsed in cosmic rays and at colliders. This imbalance in the amount of matter and anti-matter
is a long-standing and poorly understood problem in particle physics. What is known is that an essential
condition to explain the imbalance is that processes must occur in which matter can be destroyed or turned into
anti-matter. Such processes were important at the start of the universe and gave rise to today's matter-antimatter
imbalance. By searching for neutrons converting to antineutrons, the NNBAR experiment offers a unique
matter-antimatter search for such a process with a sensitivity improvement three orders of magnitude greater
than previously obtained. The observation of neutrons converting to anti-neutrons would be the first evidence
of matter changing to antimatter and would be thus of fundamental significance. In addition to addressing the
matter-antimatter imbalance, neutron--antineutron conversions tackle other longstanding and open problems in
particle physics of equally large importance. For example, it remains an enduring mystery why Nature has
provided families of two types of subatomic particles (quarks and leptons). Neutron-antineutron conversions
are predicted in theories in which quarks and leptons exist in a single theory continuing the process of unification
in physics which started with the Maxwell equations. Another mystery for which solutions predict neutronantineutron
conversions concerns the smallness of the mass of neutrinos, a neutral lepton which was considered
massless until the 1990's, compared with those of the charged leptons such as the electron. In short, at a basic
science level, the observation of neutron--antineutron conversions would be a discovery of Nobel Prize
winning significance with a direct impact on many lines of research in modern physics. 

\subsubsection{Fundamental physics with ultra-cold neutrons}
Ultra-cold neutrons play an outstanding role in addressing key questions of particle physics at the low-energy,
high-precision frontier, complementary to the high-energy frontier probed at particle accelerators. Due to the
long times for manipulation and observation in traps, UCNs are an excellent tool for precise measurements in
fundamental physics \cite{ucn1,ucn2,ucn3}. The implementation of an intense UCN source at ESS, , will boost this important field of research. Scientific
activities which will strongly profit include:

\begin{itemize}
\item { \bf Neutron electric dipole moment (EDM)}  searches are flagship experiments that have been motivating
dedicated facilities for UCN production due to their extraordinarily large discovery potential for new physics.
This experiment is famous for having ruled out more particle-physics theories than
any other experiment. As well reflected by many ambitious projects around the world, particle EDMs are among
the most promising observables to find a signal of new physics.
Calculations within the SM predict undetectably small EDMs, whereas theories beyond the SM almost
inevitably lead to much larger values. Very large
mass/energy scales can thus be probed, which may have been relevant in the evolution of the very early
Universe. The neutron is a very prominent system as it is comparably simple, without atomic shell or nuclear
structure and no net electric charge.
Ramsey's method of separated oscillatory fields, applied to a trapped ensemble of UCNs, provides highest
sensitivity to detect a neutron EDM. The still best experimental limit, |dn| < 2.9$\times$10$^{-26}$ e cm, has been obtained at
ILL by a RAL-Sussex-ILL collaboration \cite{ucn4}. There are various projects in the pipeline which were presented at
the recent workshop on Particle Physics at neutron Sources at Grenoble, aim to a level of sensitivity down to
10$^{-28}$ e cm. To be noted is that this experiment crucially depends on the amount of UCNs that can be made  \cite{ucn1,ucn2,ucn3,ucn4}.
available for trapping. Lifting the counting statistical limitations will push this experiment crucial for theoretical
particle physics to the next level of sensitivity at ESS.
\item {\bf Neutron beta decay}  studies include measurements of the neutron lifetime which is another long standing topic
of UCN physics. Despite its crucial relevance for particle physics and cosmology, its value is still not better
known than with about two per-mille accuracy, and with a significant deviation of results using different
experimental strategies. The neutron lifetime strongly impacts the abundances of light chemical elements
created during the first minutes in the evolution of our Universe \cite{ucn5}. Of similar importance for nuclear and particle
physics is that it measures the strength of semi-leptonic weak interactions and, in combination with angular
correlations in neutron decay, on one hand provides input to determine the weak axial-vector and vector
coupling constants of the nucleon, and on the other hand also enables sensitive tests of the SM, e.g., via the
unitarity of the CKM matrix~\cite{ucn1}.
\item  { \bf  Neutron gravity states for beyond-SM physics searches} is a third scientific topic which
attracts strong current interest by the scientific community and a broader public\cite{ucn6}. Quantum mechanical states of
the neutron bound by Earth's gravity to a horizontal mirror \cite{ucn7,ucn8} give access to all parameters describing Newtonian
gravitation, the mass and the distance, without disturbing Casimir or van-der-Waals effects. With their tiny nonequidistant
eigen-energies in the pico-eV range and characteristic size of a few tens of microns, these states
offer the fascinating possibility to combine tests of Newton's gravity law at short distances with the high
precision resonance spectroscopy methods of quantum mechanics. Combined with the quantum bouncer such
spectroscopic experiments also enable weak equivalence principle tests conceptually different from classical
tests, because the inertial mass $m_{i}$ and the gravitational mass $m_{g}$ enter the Schr{\"o}dinger equation. At ILL, the
qBounce collaboration has demonstrated a novel gravity resonance spectroscopy (GRS) technique that induced
resonant transitions between these quantum states by mirror vibrations \cite{ucn8}. First precision measurements sensitive
to energy changes of $\Delta$E = 10$^{-14}$ eV are described in Ref.~\cite{ucn9}. A novelty of GRS is the fact that quantum mechanical
transitions are not driven by a direct coupling of an electromagnetic charge or moment to an electromagnetic
field but by a mechanical modulator. GRS experiments offer access to search for hypothetical ``fifth" forces
provided by axions or axion-like particles, which are candidates for dark matter in the Universe whose nature
is still completely unknown. Precise GRS measurements are also sensitive to large extra dimensions and also
enable investigation of dark energy scenarios. Rates in these experiments are at the order of at best 1 UCN per
minute. Because of its large physics reach and still many orders of magnitude possible room for improvement
by technological advances, there are huge gains in sensitivity to be made with a strong UCN source at ESS.
\end{itemize}

\section{Conclusions}

ESS will start the User Program with only one moderator of exceptional brightness in place. The first 15 instruments will point to that moderator.
The start of operation of ESS will be at low power, which will be increased stepwise to finally reach its nominal  design time average power of 5 MW. 
It is planned to reach full power around 2030, by which the second moderator system that we propose could become available.
Around that date the facility would ideally be equipped with two separate
neutron sources  optimized for separate neutron characteristics, in this way ESS will offer a
versatile neutron source of outstanding performance, spanning a large neutron wavelength range without any
sacrifice in performance due to compromises to be made for a single multi-purpose source. This will enable a
plethora of multi-disciplinary activities which fit in the original plan for the ESS but offer even more
possibilities beyond, for which there is strong topical scientific demand.

\begin{acks}
The authors would like to thank  Ferenc Mezei, Yuri  Kamyshkov,  Anatolii Serebrov and Hirohiko Shimizu  for their suggestions and scientific discussions 
that helped develop the ideas in this article.

\end{acks}


\begin{thebibliography}{0}

\bibitem{1} L. Zanini et al., Design of the cold and thermal neutron moderators for the European Spallation Source, Nucl. Instr.
Meth. A {\bf 925}, 33-52 (2019).

\bibitem{2} K. Andersen, M. Bertelsen, L. Zanini, E.B. Klinkby, T. Sch${\"o}$nfeldt, P.M. Bentley, J. Saroun, Optimization of moderators
and beam extraction at the ESS, Journal of Applied Crystallography {\bf} 51, 264 (2018).
 \bibitem{jparcupgrade} https://j-parc.jp/researcher/MatLife/ja/publication/files/TS2CDR.pdf
\bibitem{snsupgrade} https://info.ornl.gov/sites/publications/Files/Pub103496.pdf
\bibitem{nnbar}M.J. Frost On behalf of the NNbar Collaboration ``The NNbar Experiment at the European Spallation Source" CPT and Lorentz Symmetry, pp. 265-267 (2017)
\bibitem{sinq} R. M. Bergmann, {it et al.}, ``Simulation Methods and Results of the SINQ Cold Neutron Source Upgrade Study"   J. Phys.: Conf. Ser. {\bf 1021}  012081 (2018) 
\bibitem{3} O. Zimmer, Superfluid-helium Ultracold Neutron sources: concepts for the European Spallation Source, Phys.
Proceedia, 51, 55 (2014).
\bibitem{4} Y. Masuda et al., Spallation UCN production for nEDM, Phys. Proceedia 51, 89 (2014).
 \bibitem{5}  J.M. Pendlebury, and J.L. Greene, Considerations for an intense source of ultracold neutrons at the European long pulse
Spallation Source, Phys. Proceedia 51, 78 (2014).
\bibitem{6}  A.R. Young et al., Spallation-driven ultracold neutron sources: concepts for a next generation source, Phys. Proceedia
51, 93 (2014).
\bibitem{7}  R. Golub, and J. Pendlebury, Phys. Lett. 53A, 133 (1975).
\bibitem{8}   R. Golub, and J. Pendlebury, Phys. Lett. A 82, 337 (1977).
\bibitem{serebrov} A.P. Serebrov, V. Liamkin et al., Development of a powerful UCN source at PNPI's WWR-M reactor, accepted for
publication in the Proceedings of the workshop ``Particle Physics at Neutron Sources", 23-25 May 2018, Grenoble.
\bibitem{12}Proceedings of the Workshop on Applications of a Very Cold Neutron Source, ANL-05/42, Argonne 2005. 
\bibitem{piegsa1}F.M. Piegsa, Phys. Rev. C {\bf } 88, 045502 (2013).
\bibitem{piegsa2} F.M. Piegsa and G. Pignol, Phys. Rev. Lett. {\bf 108}, 181801 (2012).
\bibitem{nano1} M. Teshigawara et al., Nucl. Instr. Meth. B { \bf} 450, 61-65 (2019).
 \bibitem{nano2} V. Nesvizhevsky et al., The reflection of very cold neutrons from diamond powder nanoparticles, Nucl. Instrum.
Methods Phys. Res. A { \bf 595} , 631 (2008).
\bibitem{nano3} R. Cubitt et al., Quasi-specular reflection of cold neutrons from nano-dispersed media at above-critical angles, Nucl.
Instrum. Methods Phys. Res. A { \bf 622}, 182 (2010).
\bibitem{granada} R. Granada et al., Studies on reflector materials for cold neutrons, Proceedings of the UCANS-VIII conference, 2019.
\bibitem{zimmer} O. Zimmer, Deuterated clathrate hydrates as neutron moderator and reflector in high-flux sources of cold and very cold
neutrons, Proceedings of the workshop Efficient Neutron Sources, PSI, September 2019.

\bibitem{instruments}https://www.sciencedirect.com/science/article/pii/S0168900220300097
\bibitem{ucn1}D. Dubbers, M. Schmidt, Rev. Mod. Phys. {\bf 83}, 1111 (2011).
\bibitem{ucn2} M.J. Ramsey-Musolf, S. Su, Phys. Rept. {\bf 456}, 1 (2008).
\bibitem{ucn3} H. Abele, Prog. Part. Nucl. Phys. { \bf 60}, 1 (2008).
\bibitem{ucn4} C. A. Baker, D. D. Doyle, P. Geltenbort, K. Green et al., Phys. Rev. Lett. { \bf 97}, 131801 (2006).
\bibitem{ucn5}R.H. Cyburt et al., Big bang nucleosynthesis: present status, Rev. Mod. Phys. {\bf 88}, 015004 (2016). 
\bibitem{ucn6}T. Jenke, To catch a Chameleon, Nature Physics - Measure for Measure, 13, 920 (2017).
\bibitem{ucn7} V.V. Nesvizhevsky, H.G. B${\"o}$rner, A.K. Petukhov, H. Abele et al., Nature 415, 297 (2002).
\bibitem{ucn8} T. Jenke, P. Geltenbort, H. Lemmel, H. Abele, Nature Phys. 7, 468 (2011).
\bibitem{ucn9}T. Jenke. G. Cronenberg, J. Burgd${\"o}$rfer et al., Phys. Rev. Lett. 112, 151105 (2014). 
\end{thebibliography}
\end{document}